\documentclass[a4paper]{article}
\usepackage{url}
\usepackage{INTERSPEECH2022}
\usepackage{booktabs,adjustbox,makecell,multirow}
\usepackage[colorlinks,linkcolor=red]{hyperref}

\title{The USTC-NERCSLIP Systems for the CHiME-7 DASR Challenge}
\name{Ruoyu Wang$^1$, Maokui He$^1$, Jun Du$^{1,*}$\thanks{* Corresponding author}, Hengshun Zhou$^1$, Shutong Niu$^1$, Hang Chen$^1$, Yanyan Yue$^1$, Gaobin Yang$^1$, Shilong Wu$^1$, Lei Sun$^2$, Yanhui Tu$^2$, Haitao Tang$^2$, Shuangqing Qian$^2$, Tian Gao$^2$,\\
Mengzhi Wang$^2$, Genshun Wan$^2$, Jia Pan$^2$, Jianqing Gao$^2$, Chin-Hui Lee$^3$}
\address{
  $^1$University of Science and Technology of China, China\\
  $^2$iFlytek Research, China $^3$Georgia Institute of Technology, USA}
\email{\{wangruoyu,hmk1754\}@mail.ustc.edu.cn, jundu@ustc.edu.cn}

\begin{document}

\maketitle
\begin{abstract}

This technical report details our submission system to the CHiME-7 DASR Challenge, which focuses on speaker diarization and speech recognition under complex multi-speaker scenarios. Additionally, it also evaluates the efficiency of systems in handling diverse array devices.
To address these issues, we implemented an end-to-end speaker diarization system and introduced a rectification strategy based on multi-channel spatial information. This approach significantly diminished the word error rates (WER).
In terms of recognition, we utilized publicly available pre-trained models as the foundational models to train our end-to-end speech recognition models. Our system attained a Macro-averaged diarization-attributed WER (DA-WER) of 21.01\% on the CHiME-7 evaluation set, which signifies a relative improvement of 62.04\% over the official baseline system.

\end{abstract}

\section{Introduction}

In real-world scenarios, speech signals are often accompanied by diverse environmental noises and interferences. These variations can include human voices, traffic sounds, machine noise, etc. Therefore, effectively processing and separating the speech signal of the target person in an environment with multiple sources is a challenging problem. Moreover, automatic speech recognition (ASR) in distant-talking scenarios using microphone arrays has become an integral part of our daily lives. The convenience and flexibility offered by portable devices supporting voice applications with multiple microphones have further emphasized its significance~\cite{deng2013new}. The CHiME (Computational Hearing in Multisource Environments) series challenge aims to tackle these issues and applications in multi-source speech signal processing. It motivates researchers to create novel algorithms and technologies that improve performance.

The CHiME (1-4)~\cite{barker2013pascal,vincent2013second,barker2015third} series was launched to investigate the impact of background noises in far-field scenarios and address ASR challenges in real-world applications. A common approach to enhance ASR robustness is using multi-channel speech enhancement as the front-end system. This category includes representative algorithms such as multi-channel Wiener filtering~\cite{cornelis2011performance}, blind source separation methods~\cite{comon1994independent,ono2011stable,buchner2005generalization,wang2011region}, and beamforming methods~\cite{cox1987robust,talmon2009convolutive,souden2010study}. Beamforming gained popularity in the CHiME-3 Challenge. In the CHiME-4 Challenge, the best system introduced a novel approach that combines conventional multi-channel speech enhancement with deep learning methods~\cite{tu2019an} to improve multi-channel speech recognition.

The CHiME-5~\cite{barker2018fifth} and CHiME-6~\cite{2020CHiME6} have recently provided the first large-scale corpus of real multi-talker conversational speech recorded via commercially available microphone arrays in multiple realistic homes~\cite{barker2018fifth}. 
In this challenge, the best system~\cite{sun2019a} proposed a speaker-dependent speech separation framework, exploiting advantages of both deep learning based methods and conventional preprocessing techniques.
And the CHiME-6 challenge revisits the previous CHiME-5 challenge and further considers the problem of distant multi-microphone conversational speech diarization and recognition in everyday home environments. 
In this challenge, the best system of track 1~\cite{Tu2020} proposed a space-and-speaker-aware iterative mask estimation (SSA-IME) approach to improving complex angular central Gaussian distributions (cACGMM) based beamforming in an iterative manner by leveraging upon the complementary information obtained from SSA-based regression. The best system for track 2, STC~\cite{medennikov2020stc} proposed a novel Target-Speaker Voice Activity Detection (TS-VAD) approach, which directly solves the diarization problem and allows performing GSS on top of the diarized segments.

Although the CHiME competition has achieved significant achievements in the field of multi-source speech processing, its systems are developed based on limited data and rules. Some algorithms that have won in the CHiME competition have performed well, but their generalization performance on other similar tasks is limited. Therefore, establishing a universal system in a wide range of real-world environments and providing reliable ASR performance even under adverse acoustic conditions is an important issue.

The latest CHiME-7 ~\cite{cornell2023chime} task involves using multiple recording devices for joint ASR and speaker separation in far-field environments, which may be heterogeneous. Unlike previous challenges, this challenge allows the use of external data and pre-trained models, leveraging the latest advancements in self-supervised learning and supervised learning based on DNN for speech separation and enhancement (SSE). The system evaluation includes three different scenarios (CHiME-6~\cite{2020CHiME6}, DiPCo~\cite{van2019dipco}, and Mixer 6~\cite{brandschain2010mixer}), with the goal of developing a single system that can adapt to different array geometries and use cases without any prior information while maintaining generalization capability.

This article presents our work on multi-channel processing, data augmentation, speaker diarization system, and acoustic model in the CHiME-7 Distant Automatic Speech Recognition (DASR) challenge.
Specifically: (1) We used a semi-supervised approach to utilize unlabeled data from Mixer 6 and VoxCeleb 1\&2~\cite{nagrani2017voxceleb}. 
(2) We have developed a channel selection method that adapts to various array geometries by utilizing signal to interference plus noise ratio (SINR).
(3) Recognizing the importance of spatial information in multi-channel speaker diarization systems, we developed a speaker diarization system that utilizes long-term spatial information iteratively. 
(4) By jointly fine-tuning self-supervised learning representation (SSLR), speech enhancement (SE), speaker recognition (SR), and ASR modules, we significantly improved ASR's ability to enhance target speaker performance.

\begin{figure}[t]
    \centering
    \includegraphics[width=0.45\textwidth]{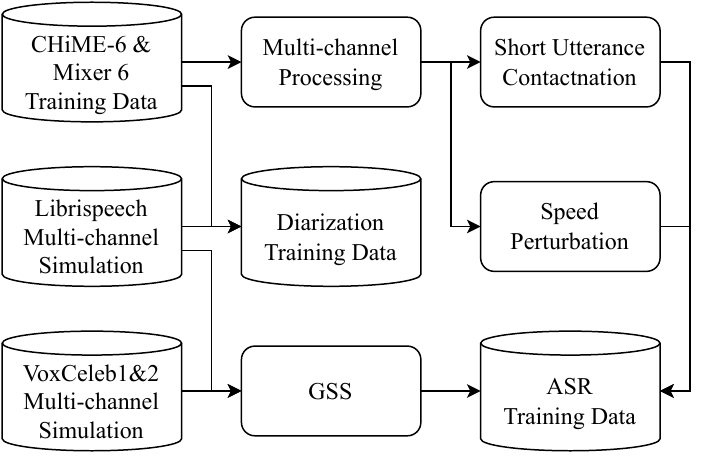}
    \caption{Data augmentation pipeline.}
    \label{fig:1.2}
\end{figure}

\section{System Description}

\subsection{Multi-channel processing}
\label{mcp}
In terms of multi-channel processing, we followed the official GPU-accelerated guided source separation (GSS) framework \cite{cornell2023chime,raj2022gpu} and made improvements in certain modules, including cross-channel synchronization, the automatic channel selection and beamforming algorithms. 

To prevent misalignments between different channels, We first calculate the lag of inter-channel correlation to perform cross-channel synchronization. In order to uniformly process multi-channel audio under different array topologies and obtain high-quality signals, we propose an automatic channel selection method based on ``virtual'' array signal to interference plus noise ratio (SINR).

It operates on two assumptions. On the one hand, the envelope variance (EV) method~\cite{cornell2023chime} can accurately rank the distances between channels and the current target speaker. Similar variance values mean that these channels may come from nearby locations. On the other hand, we assume that $N$ channels are uniformly distributed in all directions in space.

First, the values calculated using the EV method are sorted by channels.
Assuming each ``virtual'' subarray contains $K$ channels for partitioning, we obtain $\lceil N/K \rceil$ ``virtual'' subarrays.
By using the beamformed audio output from ``virtual'' subarrays, we can calculate the average SINR and sort the subarrays accordingly.
After conducting experiments on the development set, it was determined that a value of $K=5$ yielded the highest average performance across all three sets.
According to the selected subarray ratio, different versions of audio can be output. 
Specifically, we have chosen the ratio of single ``virtual'' array, front 50\% ``virtual'' array, and EV method for the first 80\% channels. 
In the beamforming section, we explore various algorithms such as minimum variance distortionless response (MVDR) beamformer and generalized eigenvalue decompositio (GEVD) beamformer. 
Furthermore, we discovered that the data processed using various algorithm settings is highly valuable during the final fusion stage.

\subsection{Data Augmentation}
\label{Data_Augmentation}

As shown in Figure~\ref{fig:1.2}, the entire training data originates from two parts, one is the CHiME-7 DASR ``Official" data \cite{officialdata}, and the other is the external data \cite{externaldata} allowed under official rules. 
We conducted a series of simulation operations based on these data to expand the data size further. 

For the first part, we directly take the manual segment boundaries of CHiME-6 as a diarization training target and multi-type GSS initialization. Since there are only transcripts and segment boundaries of subjects in the Mixer 6 training set, we first generate pseudo-labels (segment boundaries only) for interviewers by voice activity detection (VAD)-based and neural speaker diarization-based methods \cite{wu2023semisupervised}. Only the interview part of Mixer 6 is used to generate pseudo-labels and train diarization and ASR models. 

For the second part, multi-speaker (from 2 to 4 speakers) room-like multi-channel (4 channels) dialogue is simulated~\footnote{https://github.com/jsalt2020-asrdiar/jsalt2020\_simulate} using LibriSpeech~\cite{panayotov2015librispeech} and VoxCeleb 1\&2~\cite{nagrani2017voxceleb}. The simulated audio has also been added with noise extracted from non-speech segments of the CHiME-6 and Mixer 6 trainsets and music from MUSAN \cite{musan2015}. We use CHiME-6 and Mixer 6 training data and LibriSpeech multi-channel simulation data for neural speaker diarization model training.

For ASR model training, since text labels of Voxceleb 1\&2 are unusable, we first use the model trained on the other two datasets for label annotation.
The CHiME-7 DASR "Official" data was expanded with multi-type GSS to extend the audio diversity and further expanded with short utterances concatenated to long utterances and triple-speed perturbations. External data parts were produced with only a single kind of audio via the standard GSS. The specific composition of the training data is shown in Table~\ref{tab:trainingdata}.

\vspace{-0.25cm}
\begin{table}[h]
    \centering
    \caption{Composition and scale of training data.}
    \resizebox{0.47\textwidth}{!}{%
        \begin{tabular}{@{}cccccc@{}}
            \toprule
            \multirow{2}{*}{\shortstack{Datasets}} & \multirow{2}{*}{\shortstack{Original\\Duration (h)}} & \multirow{2}{*}{\shortstack{Channel\\Number}} & \multirow{2}{*}{\shortstack{Diarization\\Training (h)}} & \multirow{2}{*}{\shortstack{ASR\\Training (h)}} \\
            &&&& \\
            \midrule
            CHiME-6 & 30 & 24 & 720 & 380 \\
            Mixer 6 & 60 & 10 & 600 & 90 \\
            LibriSpeech & 960 & 4 & 3840 & 930 \\
            VoxCeleb 1\&2 & 2700 & 4 & - & 2400 \\
            \bottomrule
        \end{tabular}
    }
    \label{tab:trainingdata}
\end{table}

\subsection{Speaker Diarization}
\begin{figure}[t]
    \centering
    \includegraphics[width=0.47\textwidth]{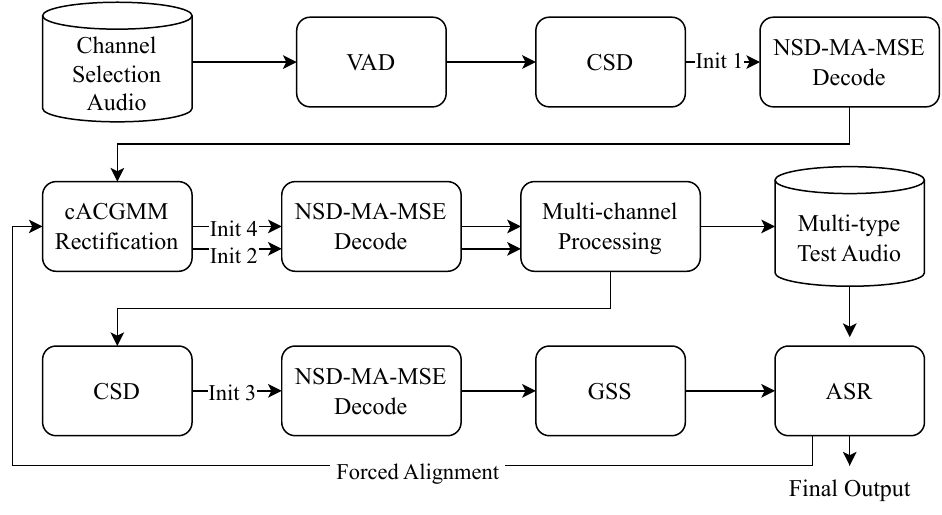}
    \caption{Multi-stage diarization inference pipeline.}
    \label{fig:1.3}
\end{figure}

The speaker diarization system is mainly based on Neural Speaker Diarization Using Memory-Aware Multi-Speaker Embedding (NSD-MA-MSE) \cite{MAMSE}. In addition to taking i-vectors as speaker embedding input in TS-VAD \cite{medennikov2020target}, Memory-Aware Multi-Speaker Embedding (MA-MSE) is concatenated to facilitate a dynamical refinement of speaker embedding to reduce a potential data mismatch between the speaker embedding extraction and the neural speaker diarization network.
Besides, we preform a sequence-to-sequence (Seq2Seq) framework in NSD-MA-MSE as in Seq2Seq-TSVAD \cite{cheng2023target}. The model is trained on both real and simulated data as mentioned in Section~\ref{Data_Augmentation}. During the inference stage, model parameters are averaged over multiple checkpoints. Speech probabilities averaged across all channels are used to generate diarization results via thresholding and post-processing. 

The NSD-MA-MSE based network also requires an initialized diarization result to generate a speaker mask matrix, where each element represents the speech/silence probability of the target speaker at each frame. With the initialization of more accurate diarization results, the decoding of the diarization system may generate more precise outcomes. This is the motivation behind our adoption of a multi-stage iterative approach. As shown in Figure~\ref{fig:1.3}, the entire diarization inference pipeline consists of multi-stage NSD-MA-MSE decoding with increasingly accurate initialized diarization inputs. 

In the first stage, the clustering-based speaker diarization (CSD) is performed on audio from EV based channel selection. Top-6 audio channels are selected to perform VAD using a baseline VAD model fine-tuned by CHiME-6 and Mixer 6 data. 

In the second stage, we perform complex Angular Central Gaussian Mixture Model (cACGMM) rectification with a window length of 120 seconds and a window shift of 60 seconds on the original audio by taking the previous NSD-MA-MSE decoding result as the initialization binary mask. 
By thresholding on the spectrum mask of cACGMM, we get the second initialized diarization result.
Through this method, we can make certain adjustments to the fusion results of single channel diarization using spatial information on all channels.

In the third stage, we perform CSD again on the official GSS-separated audio which takes the second stage NSD-MA-MSE decoding result. Our goal is to generate better clustering results by using separated audio with less noise and irrelevant speaker interference.

In the fourth stage, we once again perform cACGMM rectification on the forced alignment results. The text is generated by ASR using official GSS separated audio. HMM is obtained from kaldi tools on ASR training data.
The final diarization results are used to generate multi-type test audio using multi-channel processing in Section~\ref{mcp}.

\begin{figure}[t]
    \centering
    \includegraphics[width=0.47\textwidth]{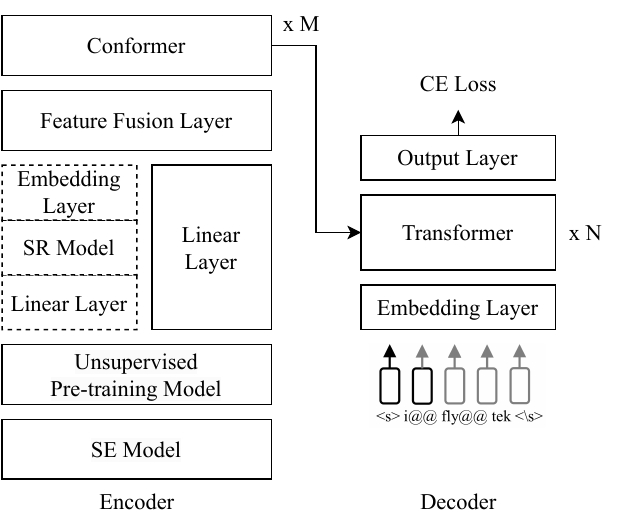}
    \caption{Joint fine-tuning framework for speaker-adaptive implicit target speaker enhancement (SAIS).}
    \label{fig:1.4}
\end{figure}
\subsection{Speech Recognition}

We propose a speaker-adaptive implicit target speaker enhancement (SAIS) approach, which is based on speaker adaptive automatic speech recognition (SA-ASR). This approach aims to efficiently optimize both SR and ASR models in order to tackle the challenge of multi-talker recognition tasks. 

The underlying principle of SA-ASR is to simply concatenate the pre-trained speaker embeddings as part of the ASR input features with acoustic features.
To optimize both speaker recognition (SR) and automatic speech recognition, we utilize self-supervised learning representation (SSLR) features from pre-trained models. These SSLR features serve as input for both SR and ASR modules, which are then fine-tuned at lower learning rates. This approach is referred to as SSLR-SA-ASR.

In order to further improve the performance of ASR for the target speaker, we optimized the SSLR, SR, and ASR modules involved in the joint fine-tuning (JFT) process. This method is referred to as speaker-adaptive implicit target speaker enhancement (SAIS) and Figure~\ref{fig:1.4} shows its framework structure. Specifically, we replace the SP layer in ECAPA-TDNN with optimization transfer (OT)~\cite{peyre2017computational} to obtain more accurate bias information for the target speaker. OT minimizes information loss by constructing mapping and cost matrices for embeddings. Additionally, we introduce hierarchy speaker-gated attention (HSGA) to effectively integrate target speaker information at each encoder layer in the ASR module. These optimizations greatly enhance ASR's ability to improve the performance of the target speaker. 
To avoid redundancy, we also attempted to change the decoder transformers' cross-attention module to memory cross-attention (MCA) module. 

Moreover, we introduced speech enhancement model (SE) as the front-end to improve the robustness, as suggested in \cite{chang2022end}.
For the SSLR extraction module, we explored the benefits of two self-supervised pre-trained models which are WavLM\cite{chen2022wavlm} and wav2vec\cite{externaldata}.
The SR module based on ECAPA-TDNN apply 512 channels and get 192 dimensions x-vector.
For the ASR module, we adopted an attention-based encoder-decoder structure.
The encoder uses a 12-layer conformer, while the decoder comprises an embedding layer, a 6-layer transformer, and an output layer.
For the SE module, we utilized the Conv-TasNet network which is pretrained on LibriSpeech and MUSAN.
In the joint fine-tuning (JFT) phase, we attempted only to update the weights with larger gradients.

The systems were trained using both real and simulated data, as discussed in Section~\ref{Data_Augmentation}. The total training data scale was approximately 3700 hours. 
The training data mentioned above exclusively consists of official CHiME-7 training data. As per the rules, researchers are allowed to rearrange the training and development sets. In accordance with this, we transferred 80\% of the utterances from the development set to the training set and applied identical data augmentation methods. This revised version of the training data amounts to 3900 hours, and all ASR models were re-trained accordingly.

\begin{table*}[h]
    \centering
    \caption{Performance comparison of different methods on CHiME-7 DEV and EVAL set (collar = 0.25 s).}
    \begin{tabular}{cccccccccc}
        \toprule
        \multirow{2}{*}{\textbf{Method}} &\multirow{2}{*}{\textbf{Set}}& \multicolumn{2}{c}{CHiME-6} & \multicolumn{2}{c}{DiPCo} & \multicolumn{2}{c}{Mixer 6} & \multicolumn{2}{c}{Macro} \\
        & & \textbf{DER} & \textbf{JER} & \textbf{DER} & \textbf{JER} & \textbf{DER} & \textbf{JER}& \textbf{DER} & \textbf{JER} \\
        \midrule
        \multirow{2}{*}{x-vectors + SC} & DEV & 40.32 & 42.31 & 24.47 & 28.97 & 15.8 & 23.07 & 26.86 & 31.45 \\
        & EVAL & 36.32 & 43.39 & 25.18 & 35.08 & 9.53 & 12.08 & 23.67 & 30.18 \\
        \multirow{2}{*}{+ NSD-MA-MSE} & DEV & 32.27 & 34.76 & 21.04 & 24.01 & 9.28 & 12.94 & 20.86 & 23.90 \\
        & EVAL & 32.09 & 37.61 & 22.78 & 31.34 & 6.21 & 7.12 & 20.36 & 25.35 \\
        \multirow{2}{*}{+ NSD-Seq2seq} & DEV & 29.93 & 33.92 & 18.22 & 22.36 & 9.85 & 13.08 & 19.33 & 23.12 \\
        & EVAL & 30.50 & 36.01 & 21.64 & 29.83 & 5.50 & 6.30 & 19.21 & 24.04 \\
        \bottomrule
    \end{tabular}
    \label{tab:method_set_performance}
\end{table*}

\begin{table*}[htb]
    \centering
    \caption{WER results of different training sets and model architectures on the CHiME-7 sub-track development set, using official GSS-generated audio.}
    \resizebox{\textwidth}{!}{%
    \begin{tabular}{cccccccccc}
        \toprule
        \textbf{ID} & \textbf{Model Architecture} & \textbf{Training Data} & \textbf{SSLR} & \textbf{SE} & \textbf{CHiME-6} & \textbf{DiPCo} & \textbf{Mixer 6} & \textbf{Macro} \\
        \midrule
        E1 & SSLR-ASR & 470h & WavLM & - & 31.66 & 34.46 & 17.86 & 27.99 \\
        E2 & SA-SSLR-ASR & 470h & WavLM & - & 31.21 & 34.19 & 17.53 & 27.64 \\
        E3 & SAIS & 470h & WavLM & - & 25.74 & 29.66 & 15.85 & 23.75 \\
        E4 & SAIS & 1400h & WavLM & - & 25.56 & 29.19 & 15.37 & 23.37 \\
        E5 & SAIS & 3800h & WavLM & - & 24.28 & 29.09 & 14.45 & 22.61 \\
        E6 & SE+SAIS & 3800h & WavLM & Frozen & 24.75 & 27.77 & 13.43 & 21.98 \\
        E7 & SE+SAIS & 3800h & WavLM+MCA & Frozen & 23.25 & 28.59 & 13.86 & 21.90 \\
        E8 & SE+SAIS & 3800h & Wav2vec & Frozen & 22.73 & 26.93 & 13.20 & 20.95 \\
        E9 & SE+SAIS & 3800h & WavLM & JFT & 22.27 & 26.94 & 12.84 & \textbf{20.68} \\
        \bottomrule
    \end{tabular}
    }
    \label{tab:experiment_results}
\end{table*}

\section{Results \& Discussion}

\subsection{Speaker Diarization}

\begin{table}[h]
    \centering
    \caption{Diarization and corresponding recognition results of four stages of iterative optimization on the CHiME-7 development set.}
    \resizebox{0.47\textwidth}{!}{%
        \begin{tabular}{ccccccccc}
            \toprule
            \multirow{2}{*}{\textbf{Stage}} & \multicolumn{2}{c}{CHiME-6} & \multicolumn{2}{c}{DiPCo} & \multicolumn{2}{c}{Mixer 6} & \multicolumn{2}{c}{Macro} \\
            & \textbf{DER} & \textbf{WER} & \textbf{DER} & \textbf{WER} & \textbf{DER} & \textbf{WER} & \textbf{DER} & \textbf{WER} \\
            \midrule
            1 & 29.93 & 33.56 & 18.22 & 35.11 & 9.85 & 12.83 & 19.33 & 27.17 \\
            2 & 27.36 & 32.78 & 16.73 & 32.01 & 9.41 & 12.41 & 17.83 & 25.73 \\
            3 & 26.53 & 30.62 & 15.83 & 30.96 & 9.17 & 12.6 & 17.18 & 24.73 \\
            4 & \textbf{25.81} & \textbf{28.61} & \textbf{15.00} & \textbf{28.63} & \textbf{8.96 }& \textbf{11.93} & \textbf{16.59} & \textbf{23.06} \\
            \bottomrule
        \end{tabular}
    }
    \label{tab:4stage}
\end{table}

Our diarization system is actually a multi-step iterative system in the CHiME-7 DASR Challenge, but for a fair comparison, we present the results of different single model systems at the first iteration in the Table~\ref{tab:method_set_performance}. Compared to NSD-MA-MSE , NSD-Seq2seq makes the macro DER drop relatively by 5.6\% on EVAL set. 

Table~\ref{tab:4stage} shows the results of our 4 stage iterative optimization initialization, decoded through NSD-MA-MSE, using official GSS and WavLM-SR-ASR acoustic models. It can be seen that with the progress of multiple stages, both DER and WER results have been gradually optimized. The average DER result of the fourth stage diarization is 16.59\%, corresponding to a WER result of 23.06\%. Compared to the first stage, there is a decrease of 13.47\% in DER and a decrease of 15.11\% in WER. 

We analyzed the error types of diarization in each stage and found that this is mainly because the introduction of spatial information in the step-by-step optimization process reduces speaker errors in the diarization results used to initialize NSD-MA-MSE decoding, resulting in more accurate decoding results. The lower DER leading to lower WER is consistent with experience.

We also tried the approach of using a deep separation model to estimate masks and iterate in our Chime-6 challenge, but it was not as effective as directly using binary masks generated from diarization results. We believe this is because cacgmm and beamforming algorithms themselves have performance bottlenecks, and in cases where diarization and ASR systems are good enough, better temporal boundary information will be more important.

\subsection{Speaker Recognition}

Table~\ref{tab:experiment_results} shows the ablation results of our structure and training data. From E1, E2, and E3, it can be seen that obtaining more effective speaker information through targeted guided acoustic models can significantly improve recognition performance in multi-speaker scenarios. The average WER of E3 decreased by 15.19\% compared to E1. 

We set up three training datasets for ablation in ascending order of scale, which are 470h, 1400h, and 3800h respectively. Please refer to Table~\ref{asrdata} for specific configurations.
\begin{table}[htb]
    \centering
    \caption{Statistics of ASR training sets.}
    \resizebox{0.47\textwidth}{!}{%
        \setlength{\tabcolsep}{4pt} 
        \fontsize{9}{11}\selectfont 
        \begin{tabular}{ccc}
            \toprule
            \textbf{Duration (h)} & \textbf{Corpus} & \textbf{Sample Scale} \\ 
            \midrule
            470 & CHiME-6 (GSS, near), Mixer 6 (near) & x3 \\
            1400 & 470 hours + LibriSpeech (simu) & x1 \\
            3800 & 1400 hours + VoxCeleb 1\&2 (simu) & x1 \\
            \bottomrule
        \end{tabular}
    }
    \label{asrdata}
\end{table}

E3, E4, and E5 demonstrated the gains brought by training data augmentation, decreasing from 23.75\% to 22.61\%, a relative decrease of 4.77\%. Finally, we added a voice enhancement module and conducted parameter freezing and joint training with other modules. The best average WER achieved by joint training of all modules is 20.68\%. 

In addition, it is noted that end-to-end speech recognition is sensitive to the length of test sentences. For too short sentences, it cannot grasp contextual information, while for too long sentences, it not only leads to errors in the GSS process but also generates some meaningless strings in ASR. Therefore, we further limit the length of test sentences. Because long-duration testing has been proven effective in many tasks, fragments from the same speaker are connected in chronological order to form sentence tests with a minimum fixed length of 10 seconds. Sentences longer than 20 seconds will be directly segmented and processed.

It is worth mentioning that, unlike many other teams in the data augmentation process, we use additional single-channel data to perform data augmentation through multi-channel and multi-speaker simulation followed by GSS. This is because we found that directly adding reverberation and vocal noise to augment single-channel data does not simulate the situation of multiple speakers speaking well. Often, it does not achieve the desired effect and requires fine adjustment of the ratio between simulated data and GSS data, sometimes even causing difficulties in model convergence.

\subsection{Overall Results}

In Table~\ref{tab:sub} and~\ref{tab:main}, we show the results of our final system. When decoding, long-time concatenation of short utterances is helpful to improve the WER metric. For different acoustic models, we used their posteriors for fusion decoding. What's more, the decoded texts of the same utterance come from multi-type test audios mentioned in Section~\ref{Data_Augmentation}, we used ROVER for the final fusion. 

In our final submission, sub-Sys1, main-Sys1, and main-Sys2 use the ASR-V1. ASR-V1 did not use data from the development set. Given that the rules allow us to re-arrange the training set and development set, we move 80\% utterances of the development set into the official training set, and perform the same data augmentation process as stated in Section~\ref{Data_Augmentation}. 
In this way, we only retrain our end-to-end ASR models (ASR-V2), and submit corresponding results as sub-Sys2 and main-Sys3. After adding dev training data, it can be observed that all subsets and macro WER have decreased. The missing DEV results are due to the use of DEV data in ASR-V2 system.

In Table~\ref{tab:main}, main-Sys1 used the best diarization result on dev in terms of WER (WER-P), and main-Sys2 used the best diarization result on dev in terms of DER (DER-P). Note that although main-Sys1 achieved better wer results than main-Sys2 on the dev set, it performed poorly on Mixer 6 subset in the eval set, whereas main-Sys2 showed satisfactory results. This to some extent indicates that a diarization system with better DER results may have stronger generalization ability.

\begin{table}[h]
    \centering
    \caption{The fusion results on CHiME-7 DASR sub-track.}
    \setlength{\tabcolsep}{14pt} 
    \begin{tabular}{cccc}
        \toprule
        \multirow{2}{*}{\textbf{System}} & \multirow{2}{*}{\textbf{Scenario}} & \multicolumn{2}{c}{\textbf{WER}} \\
        & & \textbf{DEV} & \textbf{EVAL} \\
        \midrule
        \multirow{4}{*}{Sys1} & CHiME-6  & 19.62 & 20.30 \\
             & DiPCo   & 24.12 & 20.41 \\
             & Mixer 6  & 12.16 & 14.22 \\
             & Macro   & 18.63 & 18.31 \\
        \midrule
        \multirow{4}{*}{Sys2} & CHiME-6  & -     & 19.84 \\
             & DiPCo   & -     & 19.54 \\
             & Mixer 6  & -     & 13.55 \\
             & Macro   & -     & 17.64 \\
        \bottomrule
    \end{tabular}
    \label{tab:sub}
    
\end{table}

\begin{table}[h]
    \centering
    \caption{The fusion results on CHiME-7 DASR main-track.}
    \begin{tabular}{cccccccc}
        \toprule
        \multirow{2}{*}{\textbf{System}} & \multirow{2}{*}{\textbf{Scenario}} & \multicolumn{2}{c}{\textbf{Dev}} & \multicolumn{2}{c}{\textbf{Eval}} \\
        & & \textbf{DER} & \textbf{WER} & \textbf{DER} & \textbf{WER} \\
        \midrule
        \multirow{4}{*}{Sys1} & CHiME-6 & 27.71 & 27.78 & 34.57 & 29.71 \\
         & DiPCo & 21.05 & 27.38 & 22.04 & 22.72 \\
         & Mixer 6 & 11.84 & 12.03 & 23.04 & 16.21 \\
         & Macro & 20.20 & 22.40 & 26.55 & 22.88 \\
        \midrule
        \multirow{4}{*}{Sys2} & CHiME-6 & 25.81 & 28.61 & 25.11 & 27.86 \\
         & DiPCo & 15.00 & 28.63 & 16.36 & 24.04 \\
         & Mixer 6 & 8.96 & 11.93 & 6.14 & 11.14 \\
         & Macro & 16.59 & 23.06 & 15.87 & \textbf{21.01} \\
        \midrule
        \multirow{4}{*}{Sys3} & CHiME-6 & - & - & 34.57 & 29.36 \\
         & DiPCo & - & - & 22.04 & 21.63 \\
         & Mixer 6 & - & - & 23.04 & 15.03 \\
         & Macro & - & - & 26.55 & 22.01 \\
        \bottomrule
        
    \end{tabular}
    \label{tab:main}

\end{table}

\section{Reiview \& Conclusion}

Overall, CHiME-7 Distant Automatic Speech Recognition (DASR) Challenge considered more realistic scenarios and applied more advanced technologies in the field of speech in recent years compared to previous ones. System evaluation includes three different scenarios (CHiME-6, DiPCo, and Mixer 6), which have different array geometries and acoustic environmental characteristics. Participants are required to develop a unified system without using prior information, which is a huge test for the robustness of the system.
At the same time, CHiME-7 allows the use of external data and unsupervised pre-training of large models within the rules. This brings possibilities for applying some state-of-the-art diarization methods, speech separation, and enhancement methods. 

Our team has developed a system for multi-channel processing, data augmentation, speaker separation, and acoustic modeling in the CHiME-7 DASR Challenge.
We used a semi-supervised approach to leverage the unlabeled data from Mixer 6 and VoxCeleb 1\&2 and applied GSS to generate single-channel augmented data for multi-channel multi-speaker conference scene simulation data. A channel selection method that adapts to various array geometries by utilizing signal to interference plus noise ratio (SINR) was developed. 
We have developed an iterative speaker diarization system that effectively utilizes long-term spatial information. By jointly fine-tuning self-supervised learning representation (SSLR), speech enhancement (SE), speaker recognition (SR), and ASR modules, and making modifications some of the modules, our speaker-adaptive implicit target speaker enhancement (SAIS) method significantly improved ASR's ability to enhance target speaker performance. In CHiME-7 DASR, the fusion system developed based on these methods achieved a WER of 21.01\% on the main track and won first place.

\section{Acknowledgements}

This work was supported by the National Natural Science Foundation of China under Grants No. 62171427 and No. 62001446. 

\newpage

\bibliographystyle{IEEEtran}
\bibliography{mybib}

\end{document}